\documentclass[preprint,nofootinbib]{revtex4}
\usepackage{epsfig}
\usepackage{amssymb}
\newcommand{\CC}{\Lambda}
\newcommand{\rL}{\rho_{\CC}}

\begin{document}
\begin{titlepage}

\begin{center}
{\Large \bf From inflation to dark energy through a dynamical $\CC$:
an attempt at alleviating fundamental cosmic puzzles\footnote{This essay received an honorable mention in the 2013 Essay Competition of the Gravity Research Foundation. \\$^{2}$svasil@academyofathens.gr\\$^{3}$limajas@astro.iag.usp.br\\$^{4}$sola@ecm.ub.edu}} \vskip 0.5cm {\bf Spyros Basilakos$^{a,2}$, Jos\'e Ademir Sales
Lima$^{b,3}$, Joan Sol\`a$^{c,4}$}
\end{center}

\begin{quote}
\begin{center}
$^a$Academy of Athens, Research Center for Astronomy and Applied
Mathematics, Soranou Efesiou 4, 11527, Athens, Greece\\
$^b$Departamento de Astronomia, Universidade de S\~ao Paulo, \\ Rua
do Mat\~ao 1226, 05508-900, S\~ao Paulo, SP, Brazil\\
$^c$High Energy Physics Group, Departament ECM, \\ and Institut de
Ci\`encies del Cosmos (ICC), Universitat de Barcelona,\\ Av.
Diagonal 647 E-08028 Barcelona, Catalonia, Spain
\end{center}
\end{quote}
\vspace{0.2cm}
\centerline{(Submission date: March 29, 2013)}
\vspace{0.2cm}
\centerline{\bf Abstract}
\bigskip
After decades of successful hot big-bang paradigm, Cosmology still lacks a
framework in which the early inflationary phase of the universe smoothly
matches the radiation epoch and evolves to the present `quasi' de Sitter
spacetime. No less intriguing is that the current value of the effective
vacuum energy density  is vastly smaller than the value that triggered
inflation. In this {\it Essay} we propose a new class of cosmologies
capable of overcoming, or highly alleviating, some of these acute cosmic
puzzles. Powered by a decaying vacuum energy density, the spacetime
emerges from a pure nonsingular  de Sitter vacuum stage, ``gracefully''
exits from inflation to a radiation phase followed by dark matter and
vacuum regimes, and, finally, evolves to a late time de Sitter phase.

\vspace{0.3cm}

\end{titlepage}

\pagestyle{plain} \baselineskip 0.75cm

In the standard view of cosmology, matter and space-time emerged
from a singularity and evolved through four different regimes:
inflation, radiation, dark matter and dark energy dominated eras. In
the radiation and dark matter dominated stages, the expansion of the
Universe decelerates while the inflation and dark energy eras are
accelerating phases. So far there is no 
clear cut connection between these accelerating periods. More intriguing, the substance (if any)
driving the present accelerating stage remains a complete mystery.
It has been called Dark Energy (DE), and its
formulation embodies a large number of models trying to describe its
nature and dynamical properties.

The top ten DE candidate is the cosmological constant, $\CC$, or
vacuum energy density ($\rho_{\Lambda} \equiv \Lambda/8\pi G$),
several times resurrected since Einstein introduced it 96 years ago.
Although its concept is plagued with the cosmological constant
\cite{Z67,W89} and coincidence problems \cite{CCP1,CCP2,CCP3}, most alternative DE
models present similar difficulties, and none of them can scape from
extreme fine tuning and/or insufficient fundamental
motivation\,\cite{FT1,FT2,FT3,FT4,FT5}.

It is remarkable that the Einstein field equations (plus the
Cosmological Principle) do not prevent $\CC$ to evolve with cosmic time or
a function of it. While its precise functional form has not yet been
determined, quantum field theory (QFT) in curved spacetime  singles out
the general form of the evolution of the vacuum energy density, $\rL$, as
a function of the Hubble rate. Specifically, it suggests a renormalization
group (RG) equation in which the rate of change of $\rho_{\Lambda}$ with
$H(t)$  contains only even powers of $H$ (because of the covariance of the
effective action)\,\cite{ShapSol02,BabicET02,ShapSol09,Fossil07}:
%
\begin{eqnarray}\label{seriesRLH}
\frac{d\rho_{\Lambda}}{d\ln
H^2}=\frac{1}{(4\pi)^2}\sum_{i}\left[\,a_{i}M_{i}^{2}\,H^{2}
+\,b_{i}\,H^{4}+c_{i}\frac{H^{6}}{M_{i}^{2}}\,+...\right] \,,
\end{eqnarray}
where the (dimensionless) coefficients receive loop contributions
from boson and fermion (hereafter $b$ and $f$) matter fields of
different masses $M_i$. Obviously, the expansion (\ref{seriesRLH})
converges very fast at low energies, where $H$ is rather small --
certainly much smaller than any particle mass. No other term beyond
$H^2$ (not even $H^4$) can contribute significantly on the
\textit{r.h.s.} of equation (\ref{seriesRLH}) at any stage of the
cosmological history below the GUT scale $M_{GUT}$, typically a few
order of magnitude below the Planck scale $M_P\sim 10^{19}$ GeV. But
in the very early universe (when $H$ is also close, but below, the
masses of the heavy fields $M_i\sim M_{GUT}$) the $H^4$ effects can
also be significant, whereas the terms $H^6/M_i^2$ and above are
less and less important. Integrating the above equation we arrive at
\begin{equation}\label{lambda}
\Lambda(H) = c_0 + 3\nu H^{2} + 3\alpha\,\frac{\phantom{...}H^{n+2}}{H_{I}^{n}}
\;,
\end{equation}
where $H_{I}$ is the Hubble parameter at the inflation. Here $n=2$, but we
leave it generical since the main results turn out to be independent of
$n$, as we shall see. If we banish odd powers of $H$ for the
aforementioned reasons (see however \cite{Lima94,Lima96,ML2000}) the dominant part of
the series (\ref{seriesRLH}) is expected to be naturally truncated at the
$H^4$ term\,\cite{LBS12}. The coefficients $\nu=\frac{1}{6\pi}\,
\sum_{i=f,b} c_i\frac{M_i^2}{M_P^2}$ and $\alpha=\frac{1}{12\pi}\,
\frac{H_I^2}{M_P^2}\sum_{i=f,b} b_i$ receive contributions from all the
matter particles and play the role of one-loop $\beta$-functions for the
RG running. Both coefficients are predicted to be naturally small since
$M_i^2\ll M_P^2$ for all the particles, even for the heavy fields of a
typical GUT below the Planck scale. In the case of $\nu$ an estimate
within a generic GUT is found in the range
$|\nu|=10^{-6}-10^{-3}$\,\cite{Fossil07}. Remarkably, this coefficient
can also be observationally accessed. From a joint likelihood analysis of
the recent supernovae type Ia data, the CMB shift parameter, and the BAO,
one finds that the best fit value for $\nu$ for a flat universe is
$|\nu|\lesssim{\cal O}(10^{-3})$\,\cite{BPS09,GrandeET11}, which is nicely
in accordance with the theoretical expectations.

The equations of state for the dynamical vacuum and matter fluids are
still $p_\Lambda(t)=-\rho_\Lambda(t)$ and $p=\omega \rho$ ($\omega$
constant), respectively. The overall conservation law  in the
presence of a dynamical $\CC$-term  reads
$\dot{\rho}+3(1+\omega)H\rho=-\dot{\rho_{\Lambda}}$, entailing energy
exchange between matter and vacuum. Combining this conservation
equation with (\ref{lambda}) and Friedmann's equation in flat space, $8\pi
G (\rho + \rho_{\Lambda}) = 3 H^2$, we obtain the differential evolution
law for $H(t)$:

\begin{equation}
\label{HE}
\dot H+\frac{3}{2}(1+\omega)H^2\left[1-\nu-\frac{c_0}{3H^2}-
\alpha\left(\frac{H}{H_I}\right)^{n}\right]=0 \;.
\end{equation}
Interestingly enough it admits the constant solution
$H=H_I[(1-\nu)/\alpha]^{1/n}$, corresponding to an inflationary
regime in the very early universe, i.e. when $H^2\gg c_0$. On the
other hand, at late times, when $H\ll H_I$, and for $\nu \ll 1$ we
have that $\Lambda\approx c_0$, which behaves as an effective
cosmological constant. In a nutshell, the phases of the decaying
vacuum cosmology (\ref{lambda}) are the following: {\bf (i)} the universe
starts from an unstable inflationary phase powered by the huge value
$H_I$ (presumably connected to a GUT scale near $M_P$), {\bf (ii)} it next
enters a deflationary period (with a massive production of
relativistic particles) triggering the radiation epoch, followed by
the conventional cold matter epoch, and, finally, {\bf (iii)} the vacuum
energy density effectively appears today as a slowly varying
dynamical DE, thanks to the $3\nu H^2$ term (with $|\nu|\ll 1$) in
Eq.\,(\ref{lambda}), which mildly corrects the phenomenology of the
standard $\Lambda$CDM model.

Let us first discuss the transition from an initial de Sitter stage
to the standard radiation phase ($\omega=1/3$). The Hubble function
of this model in the early universe (when $c_0$ can be neglected in
front of $H^{2}$) follows from direct integration of Eq.(\ref{HE}):
\begin{equation}\label{HS1}
 H(a)=\left(\frac{1-\nu}{\alpha}\right)^{1/n}\frac{H_I}{\left[1+D\,a^{2\,n\,(1-\nu)}\right]^{1/n}}\,.
\end{equation}
Here
$D=a_\star^{-2\,n\,(1-\nu)}\left[\frac{1-\nu}{\alpha}\left(\frac{H_I}{H_\star}\right)^n-1\right]$
is fixed from the condition $H(a_\star)\equiv H_\star$, where $a_{\star}$
is the scale factor at the transition time ($t_{\star}$) when the
inflationary period ceases. Of special physical significance are the
corresponding vacuum and radiation energy densities:
\begin{equation}\label{eq:rLa}
  \rho_\Lambda(a)=\tilde{\rho}_I\,\frac{1+\nu\,D\,a^{2n(1-\nu)}}{\left[1+D\,a^{2n(1-\nu)}\right]^{1+2/n}}\,,\ \ \  \ \
\rho_r(a)=\tilde{\rho}_I\,\frac{(1-\nu)D\,a^{2n(1-\nu)}}{\left[1+D\,a^{2n(1-\nu)}\right]^{1+2/n}}\,,
\end{equation}
where $\tilde{\rho}_I\equiv\left[(1-\nu)/\alpha\right]^{2/n}\,\rho_I$,
with $\rho_{I}=3H_{I}^{2}/8\pi G$. For $D\,a^{2n(1-\nu)}\ll1$ (the very
early universe) Eq.\,(\ref{HS1}) boils down to the aforesaid constant
value solution $H\approx H_I[(1-\nu)/\alpha]^{1/n}$. In this period the
vacuum energy density remains almost constant $\rho_\Lambda\approx
\tilde{\rho}_I$,  and the universe grows exponentially fast: $a(t)\propto
{\rm e}^{\left(\frac{1-\nu}{\alpha}\right)^{1/n}H_{I}t}$ (the primeval de
Sitter era). Right next the standard radiation dominated era emerges (see
the inner panel of Fig. 1). Thermodynamically, since the model starts as a
de Sitter spacetime, the most natural choice for the temperature is the
Gibbons-Hawking temperature \cite{GibHaw} of its event horizon. Thus the
expansion proceeds isothermally at $T_I=H_{I}/2\pi$ during the initial de
Sitter phase, with $H_I$ of the order of the  GUT scale $M_{GUT}$ near the
Planck mass  {(see \cite{LBS12} for $n=2$)}.

The outcome of the above considerations is that for $D\neq0$ the
universe starts without a singularity. Furthermore, a light pulse
beginning at $t=-\infty$ will have traveled by the cosmic time $t$ a
physical distance $d_{H}(t)= a(t)\int_{-\infty}^{t}\frac{d
t'}{a(t')}$, which diverges, thereby implying the absence of
particle horizons. As a result the local interactions may causally
homogenize the whole universe.

Remarkably, for $D\,a^{2n(1-\nu)}\gg 1$ the solution (\ref{HS1})
displays the behavior $H\sim a^{-2(1-\nu)}$ and so $a(t)\sim
t^{1/2(1-\nu)}$. As $|\nu|\ll 1$, it is obvious that $a(t)\sim
t^{1/2}$, signaling the onset of the standard radiation epoch. This
is further confirmed upon inspecting the radiation energy density in
(\ref{eq:rLa}), which decays as $\rho_r\sim a^{-4(1-\nu)}\sim
a^{-4}$. Worth noticing is that the vacuum energy density follows a
similar decay law $\rho_\Lambda\sim\nu a^{-4(1-\nu)}$, but it is
suppressed by $\rho_{\Lambda}/\rho_r\propto \nu$  (with $|\nu|\ll
1$) as compared to the radiation density. This insures that
primordial nucleosynthesis will not be harmed at all. In short, a
conventional radiation epoch is granted and a clue to the ``graceful
exit'' from the inflationary stage seems feasible.

Although we cannot provide at this point the effective action for the
model, we can at least mimic it through a scalar field ($\phi$) model for
the interacting DE \cite{SRSS,LM02,Costa08}. This can be useful for the usual
phenomenological descriptions of the DE, and can be obtained from the
correspondences
$\rho_{\rm tot} \rightarrow \rho_{\phi} =\dot{\phi}^{2}/{2} + V(\phi)$ and
$p_{\rm tot} \rightarrow p_{\phi} =\dot{\phi}^{2}/{2} - V(\phi)$ in
Friedmann equations, with the result
\begin{equation}
\label{Vz}
V(a)=\frac{3H^{2}}{8\pi G}\left( 1+\frac{\dot{H}}{3H^{2}}\right)=
\frac{\rho_I}{\alpha^{2/n}}\;\frac{1+Da^{2n}/3}{(1+Da^{2n})^{(n+2)/n}}\,,
\end{equation}
where we have now neglected the small ${\cal O}(\nu)$ corrections.
It is apparent that $V \sim \rho_I$ for $a\ll D^{-1/2n}$ (i.e.
before the transition from inflation to deflation). However, when
the transition is left well behind ($a\gg D^{-1/2n}$) the effective
potential decreases in the precise form $V(a)\sim a^{-4}$, valid for
all $n$, as it should in order to describe a radiation dominated
universe -- independently of the power $n$. This result
corroborates, in the scalar field language, the correct transition
to the radiation dominated epoch.

\begin{figure*}
\centerline{\epsfig{file=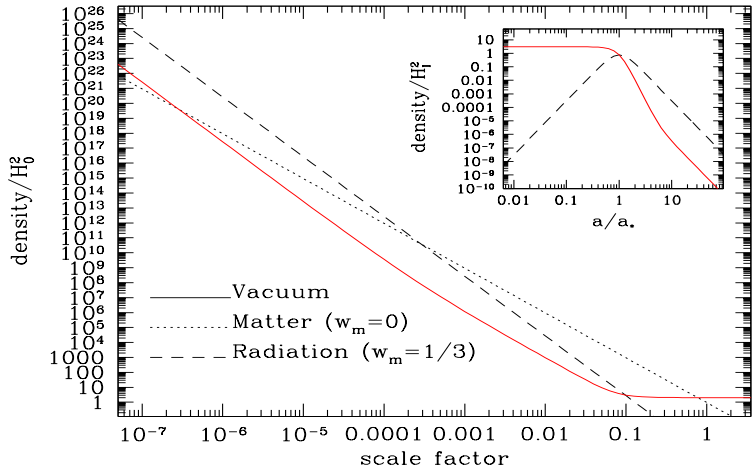,width=120mm, height=70mm} \hskip 0.1in}
\caption{{\bf Outer Plot:} The evolution of the radiation,
non-relativistic matter and vacuum energy densities, for the unified
vacuum model (\ref{lambda}) with  $n=2$ in units of $H^{2}_{0}$. Note that
similar behavior is found also $\forall n$. The curves shown are:
radiation (dashed line), non-relativistic matter (dotted line) and vacuum
(solid line, in red). To produce the lines we used $\nu=10^{-3}$,
$\Omega^{0}_{m}=0.3175$, 
$\Omega^{0}_{R}=(1+0.227N_{v})\,\Omega^{0}_{\gamma}$,
$\Omega^{0}_{\Lambda}=1-\Omega^{0}_{m}-\Omega^{0}_{R}$,
$(N_{v},\Omega^{0}_{\gamma},h)\simeq (3.04,2.47\times 10^{-5}h^{-2},0.6711)$
(cf. Ade {\it et al.} \cite{Ade13}, {\it Planck results}), and set $\alpha=1$ and
$D=1/a_{*}^{3(1-\nu)(1+\omega)}$. {\bf Inner Plot:} the primeval vacuum
epoch (inflationary period) into the FLRW radiation epoch. Same notation
for curves as before, although the densities are now normalized with
respect to $H_I^2$ and the scale factor with respect to $a_{*}$. For
convenience we used $8\pi G=1$ units in the plots.}\label{fig1}
\end{figure*}

Finally, we briefly analyze the expanding universe at times after
recombination, therefore consisting of dust ($\omega=0$) plus the running
vacuum fluid described by Eq.(\ref{lambda}) with $H\ll H_I$. In this case
the $H^{n+2}$ term ($n>1$) is completely negligible compared to $H^2$ and
we have $\Lambda(H)=\Lambda_0+3\,\nu\,(H^2-H_0^2)$, with $\Lambda_0\equiv
c_0+3\nu\,H_0^2$. Obviously, $c_0$ plays an essential role to determine
the value of $\Lambda$, and since $|\nu|={\cal O}(10^{-3})$ the $H^2$
dependence gives some remnant dynamics even today. Trading the cosmic time
for the scale factor and using the redshift variable $1+z=1/a$ with the
boundary condition $H(z=0)=H_0$, one finds the solution of Eq.(\ref{HE})
for the late stages:
\begin{equation}
\label{Hz}
 H^{2}(z)= \frac{H_0^2}{1-\nu} \left[(1-\Omega_{\Lambda}^0)(1+z)^{3(1-\nu)}+
\Omega_{\Lambda}^0-\nu \right]\,,
\end{equation}
where
$\Omega_{\Lambda}^0=\Lambda_{0}/3H_0^{2}=8\pi\,G\,\rho_{\Lambda}^0/3H_0^{2}$.
The corresponding matter and vacuum energy densities read:
$\rho_{m}(z)=\rho_{m}^{0}(1+z)^{3(1-\nu)}$ and
$\rho_{\Lambda}(z)=\rho^{0}_{\Lambda}+\frac{\nu\,\rho_{m}^{0}}{1-\nu}\left[(1+z)^{3(1-\nu)}-1\right]$.
They deviate from the $\Lambda$CDM, but for $\nu\to 0$ they retrieve
their standard forms (in particular, $\rL$ becomes constant).
Recalling that $|\nu|\ll 1$, the model is almost indistinguishable
from the concordance $\Lambda$CDM, except for its mild vacuum
dynamical behavior which leads to an effective equation of state
that can mimic quintessence or phantom
energy\,\cite{SolaStefancic05,SolaStefancic06}. {This also means that structure
formation after recombination evolves like in the $\CC$CDM model.}
At very late times, $H$ becomes constant again: $H\approx
H_0\,\sqrt{(\Omega_{\Lambda}-\nu)/(1-\nu)}$, hence opening up a new
pure de Sitter phase. As an example, in Fig. 1 we display the case
$n=2$, with the mentioned details for the early (inner plot) and
intermediate/late (outer plot) stages  of the cosmic evolution. For
$z\le 10$ (or $a\ge 0.1$) the vacuum energy density appears
virtually frozen to its nominal value, $\rho_{\Lambda}\approx
\rho^{0}_{\Lambda}$, close to the matter density.

The upshot is a unified vacuum picture, spanning the entire history of the
universe and deviating at present only very mildly from the observed
$\Lambda$CDM behavior. For any power $n>1$ in (\ref{lambda}), the value of
$\Lambda$ at the early de Sitter phase is $\Lambda_I \sim H^{2}_I$ while
at present $\Lambda_0 \sim H^{2}_{0}$.  For $H_I$  near the Planck
scale, the correct ratio $\Lambda_I/\Lambda_0 \sim 10^{122}$ ensues.




\begin{thebibliography}{99}




\bibitem{Z67} Ya. B. Zeldovich, JETP Lett. 6, 316 (1967).

\bibitem{W89} S. Weinberg, {\em Rev. Mod. Phys.} {61}, 1 (1989).

\bibitem{CCP1}  P.~J.~E. Peebles and B.
Ratra, {\em Rev. Mod. Phys.} 75, 559 (2003).

\bibitem{CCP2} T.~Padmanabhan, {\em Phys. Rept.} 380, 235 (2003). 

\bibitem{CCP3} J. A. S. Lima, {\em Braz. J. Phys.} {34}, 194 (2004), {\bf astro-ph/0402109}  

\bibitem{FT1}  V. Sahni and A. A. Starobinsky, {\em Int. J. Mod. Phys.} D9, 373
(2000).  

\bibitem{FT2} C. Wetterich,  {\em Phys. Lett.} B594, 17 (2004).

\bibitem{FT3} J. F. Jesus {\it et al.}, {\em Phys. Rev.} D78, 063514
(2008). 

\bibitem{FT4} S. Basilakos, M. Plionis and J. A. S. Lima, {\em Phys. Rev.}
D82,  083517 (2010).

\bibitem{FT5} S. Basilakos, F. Bauer, J. Sol{\`a}, {\em JCAP}
1201, 050 (2012).

\bibitem{ShapSol02} I.~L. Shapiro and J.~Sol{\`a}, {\em Phys. Lett.} B475,
    236, (2000);  {\em JHEP} 02, 006 (2002).

\bibitem{BabicET02} A.~Babi{\'c}, B.~Guberina, R.~Horvat and
    H.~\v{S}tefan\v{c}i{\'c}, {\em Phys. Rev.} D65, 085002 (2002).


\bibitem{ShapSol09} I. L. Shapiro and J. Sol{\`a}, {\em Phys. Lett.}
    B682, 105 (2009).

\bibitem{Fossil07} J.~Sol{\`a}, {\em J. of Phys.} A41, 164066 (2008).

\bibitem{Lima94} J. A. S. Lima, J. M. F. Maia, {\em Phys. Rev.} D49,  5597 (1994).

\bibitem{Lima96} J. A. S. Lima and M. Trodden, {\em Phys. Rev.} D53,  4280 (1996).

\bibitem{ML2000} J. M. F. Maia, \textit{Some Applications of Scalar Fields in
    Cosmology}, PhD thesis (in Portuguese),  S\~ao Paulo University
    (2000).

\bibitem{LBS12} J. A. S. Lima, S. Basilakos and J. Sol\`a,  {\em MNRAS} 431, 923 (2013), {\bf arXiv:1209.2802}.

\bibitem{BPS09} S.~Basilakos, M.~Plionis and J.~Sol{\`a}, {\em Phys. Rev.} D80, 083511 (2009).

\bibitem{GrandeET11} J.~Grande, J.~Sol{\`a}, S.~Basilakos and M.~Plionis, {\em JCAP} 08, 007 (2011).

\bibitem{GibHaw} G. Gibbons and S. W. Hawking, {\em Phys. Rev.} D15, 2738 (1977).

\bibitem{SRSS} T. D. Saini, S. Raychaudhury, S. Sahni and A. A. Starobinsky, {\em Phys. Rev. Lett.} 85, 1162 (2000). 

\bibitem{LM02} J. M. F. Maia and J. A. S. Lima, {\em Phys. Rev.} D65, 083513 (2002).

\bibitem{Costa08} F.~E.~M.~Costa, J.~S.~Alcaniz and J.~M.~F.~Maia, {\em Phys. Rev.} D77, 083516 (2008).

\bibitem{Ade13} P. A. R. Ade {\it et al.}, ``Planck Collaboration XVI. Cosmological Parameters'', {\bf arXiv:1303.5076}

\bibitem{SolaStefancic05} J.~Sol{\`a} and H.~\v{S}tefan\v{c}i{\'c}, {\em Phys. Lett. } B62, 147 (2005).

\bibitem{SolaStefancic06} J.~Sol{\`a} and H.~\v{S}tefan\v{c}i{\'c}, {\em Mod. Phys. Lett.} A21, 479 (2006).

\end{thebibliography}
\end{document}